\documentclass[aps,twocolumn,preprintnumbers,amsmath,amssymb,superscriptaddress,floatfix,nofootinbib]{revtex4-1}
\usepackage{graphicx} 
\usepackage{amsmath,amssymb}
\usepackage{subfigure}
\usepackage{multirow} 
\usepackage[vcentermath]{youngtab}
\usepackage[colorlinks,linkcolor=blue]{hyperref}
\usepackage{newtxtext, newtxmath}
\usepackage{braket,bm}
\usepackage{enumitem}
\allowdisplaybreaks

\begin{document}

\title{Hidden-charm pentaquark states $qqqc\bar{c}$ $(q = u,d)$ in the chiral SU(3) quark model}

\author{Du Wang}
\affiliation{School of Nuclear Science and Technology, University of Chinese Academy of Sciences, Beijing 101408, China}

\author{Wen-Ling Wang}
\email[Email: ]{wangwenling@buaa.edu.cn}
\affiliation{School of Physics, Beihang University, Beijing 100191, China}

\author{Fei Huang}
\email[Email: ]{huangfei@ucas.ac.cn}
\affiliation{School of Nuclear Science and Technology, University of Chinese Academy of Sciences, Beijing 101408, China}

\date{\today}

\begin{abstract}
In this work, we systematically calculate the spectrum of hidden-charm pentaquark states $qqqc\bar{c}$  $(q = u,d)$ in the chiral SU(3) quark model, which has been quite successful in reproducing consistently the energies of octet and decuplet baryon ground states, the binding energy of deuteron, and the nucleon-nucleon ($NN$) scattering phase shifts and mixing parameters for partial waves with total angular momentum up to $J=6$. The Hamiltonian contains the one-gluon-exchange (OGE) potential, the Goldstone-boson-exchange (GBE) potential, the confinement potential, and the kinetic energy of the system. We solve the Schr\"odinger equation by use of the variational method. It is found that the masses of all the experimentally observed $P_c(4312)$, $P_c(4380)$, $P_c(4440)$, and $P_c(4457)$ states are much overestimated, indicating that these states are not compact pentaquark states in the chiral SU(3) quark model. All other $qqqc\bar{c}$ $(q = u,d)$ states are found to lie much above the corresponding baryon-meson thresholds, and thus are not suggested as stable pentaquark states due to their fall-apart decays. A detailed comparison of the results with those obtained in the OGE model and the chromomagnetic interaction (CMI) model is further given. 
\end{abstract}

\maketitle

\section{Introduction}

Before 2003, most people believe that hadrons can be divided into two categories, i.e. meson consists of a pair of quark and antiquark ($q\bar{q}$) and baryon consists of three quarks ($qqq$). This view underwent a change in 2003, due to the discovery of the famous $X(3872)$ by the Belle Collaboration \cite{Belle:2003nnu}. Different from traditional meson ($q\bar{q}$) and baryon ($qqq$), $X(3872)$ probably consists of configuration $c\bar{c}q\bar{q}$ $(q=u,d)$. Later, the subsequent observation of a series of multiquark candidates like $Y(3940)$ \cite{Belle:2004lle}, $Z^+(4430)$ \cite{Belle:2007hrb}, $X(3823)$ \cite{Belle:2013ewt}, etc., has enhanced our belief of the existence of multiquark states. Until now, dozens of exotic hadrons states were reported by different collaborations and have aroused significant research interest. Recent experimental and theoretical status on exotic states can be found in Refs.~\cite{Hosaka:2016pey,Ali:2017jda,Guo:2017jvc,Esposito:2016noz,Lebed:2016hpi,Richard:2016eis,Chen:2016qju,Liu:2019zoy,Brambilla:2019esw}. 

In 2015, the LHCb Collaboration reported two exotic states, $P_c(4380)$ and $P_c(4450)$, in the $J/\psi p$ invariant mass spectrum of the $\Lambda_b^0\to J/\psi K^-p$ decay process \cite{LHCb:2015yax}. Four years later, in 2019, the LHCb Collaboration claimed that the $P_c(4450)$ should actually be resolved into two narrow states,  $P_c(4440)$ and $P_c(4457)$, and a new state $P_c(4312)$ was reported at the same time \cite{LHCb:2019kea}. The masses and widths of these observed four $P_c$ states are
\begin{align*}
\begin{array}{rrrl}
P_{c}(4380)^{+}: && M\!\! &= 4380 \pm 8 \pm 29 ~\mathrm{MeV}, \\[5pt]
&& \Gamma \!\! & = 215 \pm 18 \pm 86 ~\mathrm{MeV}, \\[8pt]
P_{c}(4312)^{+}: && M\!\!  &= 4311.9 \pm 0.7_{-0.6}^{+6.8} ~\mathrm{MeV}, \\[5pt]
&& \Gamma \!\!  &= 9.8 \pm 2.7_{-4.5}^{+3.7} ~\mathrm{MeV}, \\[8pt]
P_{c}(4440)^{+}: && M \!\! &= 4440.3 \pm 1.3_{-4.7}^{+4.1} ~\mathrm{MeV}, \\[5pt]
&& \Gamma \!\! &= 20.6 \pm 4.9_{-10.1}^{+8.7} ~\mathrm{MeV}, \\[8pt]
P_{c}(4457)^{+}: && M \!\!  &= 4457.3 \pm 0.6_{-1.7}^{+4.1} ~\mathrm{MeV}, \\[5pt]
&&\Gamma \!\! &= 6.4 \pm 2.0_{-1.9}^{+5.7} ~\mathrm{MeV}.
\end{array}
\end{align*}
These $P_c$ states were suggested as candidates of hidden-charm pentaquark states. 

The discovery of the $P_c$ states has sparked researcher's great interest. The inner structure of these pentaquark states got a lot of attention and was explored with various methods, e.g., quark models \cite{Weng:2019ynv,Deng:2022vkv,Li:2023aui,Hiyama:2018ukv,Cheng:2019obk}, QCD sum rules \cite{Chen:2015moa,Chen:2019bip,Chen:2016otp,Zhang:2019xtu}, and the one-boson-exchange (OBE) model \cite{Chen:2019asm}, et al. Unfortunately, the conclusions drawn from different works are not consistent yet. Within the quark model, Refs.~\cite{Weng:2019ynv,Deng:2022vkv,Li:2023aui} supported the compact pentaquark explanation of the experimentally observed $P_c$ states. However, in Ref.~\cite{Hiyama:2018ukv}, Hiyama {\it et al.} found two sharp resonant states at $4690$ MeV and $4920$ MeV, while no resonance was found in the energy region of $4300$-$4500$ MeV, and thus, the $P_c$ states were not regarded as compacted pentaquark states in their quark model calculation. In Ref.~\cite{Cheng:2019obk}, Chen {\it et al.} claimed that the $P_c$ states can be explained in both compact pentaquark picture and molecule picture in their chromomagnetic interaction (CMI) quark model. Similarly, contradictory conclusions were also drawn in studies performed by use of QCD sum rules \cite{Chen:2015moa,Chen:2019bip,Chen:2016otp,Zhang:2019xtu}. In Refs.~\cite{Chen:2015moa,Chen:2019bip,Chen:2016otp}, the authors reported that the molecular state structure is appropriate to explain the observed $P_c$ states. While in Ref.~\cite{Zhang:2019xtu}, Wang {\it et al.} explained the structures of the $P_c$ states in a diquark–diquark–antiquark picture. In the OBE model study of Ref.~\cite{Chen:2019asm}, Chen {\it et al.} claimed that the $P_c(4312), P_c(4440)$, and $P_c(4457)$ states correspond to loosely bound $\Sigma_{c} \bar{D}$, $\Sigma_{c} \bar{D}^{*}$, and $\Sigma_{c} \bar{D}^{*}$ molecular states, respectively. 

It is seen that although a lot of theoretical works have already been devoted to study the $P_c$ states, conclusions about the structures of these states drawn by different theoretical works were still inconclusive, regardless of whether these works were completed with the same or different theoretical methods. This poor situation makes one realize the importance and necessity of further independent analysis with reliable models in order to get a better understanding of the properties of the hidden-charm pentaquark states. 

In our previous work of Ref.~\cite{Huang:2018rpb}, we have successfully described the energies of octet and decuplet baryon ground states, the binding energy of deuteron, the nucleon-nucleon ($NN$) scattering phase shifts and mixing parameters for partial waves with angular momentum up to $J=6$ within a chiral SU(3) quark model. The Hamiltonian includes the kinetic energy, the one-gluon-exchange (OGE) potential, the confinement potential, and the Goldstone-boson-exchange (GBE) potential generated by the coupling of quark and chiral fields. It is worth mentioning that the work of Ref.~\cite{Huang:2018rpb} was, as far as we know, the first and so far the only work that reproduces the energies of octet and decuplet baryon ground states and the experimental data of $NN$ scattering in a quite consistent way. It solved the problem that the wave functions selected for single baryons are not consistent with those for two-baryon states in resonating group method (RGM) study of baryon-baryon interactions in constituent quark models. 

In the present work, we further extend the chiral SU(3) quark model employed in Ref.~\cite{Huang:2018rpb} to explore the mass spectra of the $qqqc\bar{c}$ $(q = u,d)$ hidden-charm pentaquark systems. The interactions between a pair of light quarks $qq$ $(q = u,d)$ are taken from Ref.~\cite{Huang:2018rpb}. The interactions associated with charm quark and antiquark consist of the OGE potential and the confinement potential, and the corresponding parameters are fixed by a fit to the masses of known charmed baryons and mesons. The total wave functions of the $qqqc\bar{c}$ $(q = u,d)$ pentaquark systems are constructed as combinations of the wave functions in color, flavor, spin, and orbit spaces under the constraints of the Pauli principle.  The spacial trial wave functions are chosen as Gaussian functions. The masses and eigenvectors for the $qqqc\bar{c}$ $(q = u,d)$ hidden-charm pentaquark states are obtained by solving the Schr\"odinger equation via the variational method. 

The paper is organized as follow. In Sec.~\ref{sec:wavef_Hamil}, we construct the wave functions for the $qqqc\bar{c}$ $(q=u,d)$ systems and introduce the Hamiltonian employed in the chiral SU(3) quark model. In Sec.~\ref{sec:results}, we present the numeric results of the mass spectra of the $qqqc\bar{c}$ $(q=u,d)$ states and discuss the difference compared with the results from OGE model and CMI model. In Sec.~\ref{sec:summary}, we give a brief summary.

\section{Wave function and Hamiltonian}\label{sec:wavef_Hamil}

\subsection{Wave function}

We introduce the following Jacobi coordinates for the $qqqc\bar{c}$ $(q=u,d)$ systems:
\begin{align}
\bm{\xi}_1 &= \bm{r}_{1} - \bm{r}_{2},   \\[6pt]
\bm{\xi}_2 &= \frac{\bm{r}_{1} + \bm{r}_{2}}{2} - \bm{r}_{3},   \\[6pt]
\bm{\xi}_3 &= \bm{r}_{4} - \bm{r}_{5},   \\[6pt]
\bm{\xi}_4 &= \frac{\bm{r}_{1} + \bm{r}_{2} + \bm{r}_{3}}{3} - \frac{\bm{r}_{4} + \bm{r}_{5}}{2},   \label{Eq:Jacobi-xi4}
\end{align}
where $\bm{r}_1$, $\bm{r}_2$, and $\bm{r}_3$ are coordinates of three light quarks in $qqqc\bar{c}$ $(q=u,d)$, and $\bm{r}_4$ and $\bm{r}_5$ are coordinates of charm quark and antiquark, respectively. The spacial wave function of the $qqqc\bar{c}$ $(q=u,d)$ pentaquark systems are constructed as 
\begin{equation}
\phi_{\rm space} \,=\, \prod_{i=1}^4 \, \left(\frac{2\nu_i}{\pi}\right)^{3/4} e^{-\nu_i \bm{\xi}_i^2} \, ,  \label{eq:spacial}
\end{equation}
where $\nu_i$ $(i=1-4)$ are Gaussian width parameters which will be determined by the variational method. Note that the relation $\nu_2 \equiv 4\nu_1/3$ is fixed during the variational procedure to ensure that the spacial wave functions of three light quarks are symmetric.

As the spacial wave function constructed in Eq.~(\ref{eq:spacial}) is already symmetric for three light quarks, the total wave function of the $qqqc\bar{c}$ $(q=u,d)$ system can be constructed under the constraints that the color-spin-flavor wave function is antisymmetric for three light quarks and the whole five quark system is colorless. In view of this, the most convenient way to construct the total wave function of the $qqqc\bar{c}$ system is to construct the total wave functions for the $qqq$ cluster and $c\bar{c}$ cluster separately and then combine them together, e.g.
\begin{equation}
\psi_{qqqc\bar{c}} = \psi_{qqq} \, \psi_{c\bar{c}},
\end{equation}
where $\psi_{qqq}$ is totally antisymmetric and $\psi_{qqqc\bar{c}}$ is colorless. The color wave function can be constructed in the following two different ways:
\begin{align}
\left | {\bf{1}}_c {\bf{1}}_c \right \rangle & \equiv \left [ {\left(qqq\right)}^{{\bf{1}}_c} {\left(c\bar{c}\right)}^{{\bf{1}}_c} \right ] ^{{\bf{1}}_c},  \label{eq:color-1c1c} \\[4pt]
\left | {\bf{8}}_c {\bf{8}}_c \right \rangle & \equiv \left [ {\left(qqq\right)}^{{\bf{8}}_c} {\left(c\bar{c}\right)}^{{\bf{8}}_c} \right ] ^{{\bf{1}}_c},  \label{eq:color-8c8c}
\end{align}
where ${\bf{1}}_c$ and ${\bf{8}}_c$ denote the irreducible representations of the color SU(3) group, respectively. For $\left | {\bf{1}}_c {\bf{1}}_c \right \rangle $ color configuration, the spin-flavor wave function for three light quarks should be symmetric as their color wave function is already antisymmetric. The symmetric spin-flavor wave function for three light quarks can be denoted as either $\{ qqq \}_{3/2}$ or $(qqq)_{1/2}$, where the curly brace and the parentheses represent that the three light quarks have isospin $3/2$ and $1/2$, corresponding to a complete symmetry and mixed symmetry in the flavor space, respectively, and the subscripts denote the spin of these three light quarks. For $\left | {\bf{8}}_c {\bf{8}}_c \right \rangle $ color configuration, the spin-flavor wave function for three light quarks should have mixed symmetry under a permutation of any pair of quarks, and it can be constructed as one of $\{ qqq \}_{1/2}$, $(qqq)_{3/2}$, and $(qqq)_{1/2}$, which results in a totally antisymmetric wave function when combined  with the color wave function. For $c\bar{c}$, there is no further constraints for its quantum numbers, and its spin can be either $0$ or $1$.

\begin{table}[tbp]
\caption{ Configurations of $S$-wave $qqqc\bar{c}$ $(q=u,d)$ systems. The superscripts and subscripts represent the color SU(3) representations and the spin quantum numbers, respectively. The parentheses $(~)$ and the curly brace $\{~\}$ for the $qqq$ cluster represent that the isospin of these three light quarks are 1/2 and 3/2, respectively. }
\begin{tabular*}{\columnwidth}{@{\extracolsep\fill}ccc}
\hline\hline
  $IJ^P$ & \multicolumn{2}{c}{Configuration}  \\ 
\hline
$\frac{1}{2}\frac{1}{2}^-$ & $\left\{\left(qqq\right)^{{\bf{1}}_c}_{1/2}\left[c\bar{c}\right]^{{\bf{1}}_c}_{0}\right\}^{{\bf{1}_c}}_{1/2}$ & $ \left\{\left(qqq\right)^{{\bf{1}}_c}_{1/2}\left[c\bar{c}\right]^{{\bf{1}}_c}_{1}\right\}^{{\bf{1}}_c}_{1/2}$ \vspace{0.5em} \\
& $\left\{\left(qqq\right)^{{\bf{8}}_c}_{3/2}\left[c\bar{c}\right]^{{\bf{8}}_c}_{1}\right\}^{{\bf{1}}_c}_{1/2}$ & $\left\{\left(qqq\right)^{{\bf{8}}_c}_{1/2}\left[c\bar{c}\right]^{{\bf{8}}_c}_{0}\right\}^{{\bf{1}}_c}_{1/2}$ \vspace{0.5em} \\
& $\left\{\left(qqq\right)^{{\bf{8}}_c}_{1/2}\left[c\bar{c}\right]^{{\bf{8}}_c}_{1}\right\}^{{\bf{1}}_c}_{1/2}$\vspace{1em} \\
$\frac{1}{2}\frac{3}{2}^-$  & $\left\{\left(qqq\right)^{{\bf{1}}_c}_{1/2}\left[c\bar{c}\right]^{{\bf{1}}_c}_{1}\right\}^{{\bf{1}}_c}_{3/2}$ & $\left\{\left(qqq\right)^{{\bf{8}}_c}_{3/2}\left[c\bar{c}\right]^{{\bf{8}}_c}_{0}\right\}^{{\bf{1}}_c}_{3/2}$ \vspace{0.5em} \\
& $\left\{\left(qqq\right)^{{\bf{8}}_c}_{3/2}\left[c\bar{c}\right]^{{\bf{8}}_c}_{1}\right\}^{{\bf{1}}_c}_{3/2}$ & $\left\{\left(qqq\right)^{{\bf{8}}_c}_{1/2}\left[c\bar{c}\right]^{{\bf{8}}_c}_{1}\right\}^{{\bf{1}}_c}_{3/2}$\vspace{1em} \\
$\frac{1}{2}\frac{5}{2}^-$ & $\left\{\left(qqq\right)^{{\bf{8}}_c}_{3/2}\left[c\bar{c}\right]^{{\bf{8}}_c}_{1}\right\}^{{\bf{1}}_c}_{5/2}$ \vspace{1em} \\
$\frac{3}{2}\frac{1}{2}^-$ & $\left\{\left\{qqq\right\}^{{\bf{1}}_c}_{3/2}\left[c\bar{c}\right]^{{\bf{1}}_c}_{1}\right\}^{{\bf{1}}_c}_{1/2}$ & $\left\{\left\{qqq\right\}^{{\bf{8}}_c}_{1/2}\left[c\bar{c}\right]^{{\bf{8}}_c}_{0}\right\}^{{\bf{1}}_c}_{1/2}$ \vspace{0.5em}  \\
& $\left\{\left\{qqq\right\}^{{\bf{8}}_c}_{1/2}\left[c\bar{c}\right]^{{\bf{8}}_c}_{1}\right\}^{{\bf{1}}_c}_{1/2}$  \vspace{1em}  \\
$\frac{3}{2}\frac{3}{2}^-$ & $\left\{\left\{qqq\right\}^{{\bf{1}}_c}_{3/2}\left[c\bar{c}\right]^{{\bf{1}}_c}_{0}\right\}^{{\bf{1}}_c}_{3/2}$ & $\left\{\left\{qqq\right\}^{{\bf{1}}_c}_{3/2}\left[c\bar{c}\right]^{{\bf{1}}_c}_{1}\right\}^{{\bf{1}}_c}_{3/2}$ \vspace{0.5em} \\
& $\left\{\left\{qqq\right\}^{{\bf{8}}_c}_{1/2}\left[c\bar{c}\right]^{{\bf{8}}_c}_{1}\right\}^{{\bf{1}}_c}_{3/2}$\vspace{1em}\\
$\frac{3}{2}\frac{5}{2}^-$ & $\left\{\left\{qqq\right\}^{{\bf{1}}_c}_{3/2}\left[c\bar{c}\right]^{{\bf{1}}_c}_{1}\right\}^{{\bf{1}}_c}_{5/2}$ \\
\hline\hline
\end{tabular*}  \label{tab:wavef}
\end{table}

All configurations for the $S$-wave $qqqc\bar{c}$ $(q=u,d)$ systems are listed in Table~\ref{tab:wavef}. There, the superscripts and subscripts represent the color SU(3) representations and the spin quantum numbers, respectively. The parentheses $(~)$ and the curly brace $\{~\}$ for the $qqq$ cluster represent that the isospin of these three light quarks are 1/2 and 3/2, respectively.

\subsection{Hamiltonian}

The Hamiltonian of the hidden-charm pentaquark $qqqc\bar{c}$ $(q = u,d)$ system consists of the masses of constituent quarks, the kinetic energy, and the potential between constituent quarks, e.g., 
\begin{equation}  \label{eq:Hamiltonian}
H = \sum_{i=1}^5 \left( m_i + T_i \right ) - T_G + \sum_{1=i<j}^5 \left( V_{ij}^{\text{conf}} + V_{ij}^{\text{OGE}} + V_{ij}^{\text{GBE}} \right),
\end{equation}
where $m_i$ and $T_i$ represent the mass and kinetic energy of the $i$th constituent quark, respectively. $T_G$ is the kinetic energy of the center-of-mass of the $qqqc\bar{c}$ system, 
\begin{equation}
T_i = \frac{{\boldsymbol{p}}_i^2}{2m_i}, \qquad T_G = \frac{ \left(\sum_{i=1}^5 \boldsymbol{p}_i\right)^2  }{ 2 \sum_{i=1}^5 m_i },
\end{equation}
with $\boldsymbol{p}_i$ being the three-momentum of the $i$th constituent quark. 
The potential between a pair of constituent quarks consists of three parts: the phenomenological confinement potential $V_{ij}^{\text{conf}}$, the OGE potential $ V_{ij}^{\text{OGE}}$, and the GBE potential $V_{ij}^{\text{GBE}}$. Note that the GBE potential exists only between a pair of light quarks in the chiral SU(3) quark model \cite{Huang:2018rpb}. 

The confinement potential $V_{ij}^{\text{conf}}$ phenomenally describes the long-range non-perturbative QCD effects. In the present work, we adopt the linear type confinement potential, 
\begin{equation}
V_{ij}^{\text{conf}} = - \boldsymbol{\lambda}_i^c \cdot \boldsymbol{\lambda}_j^c \left( a_{ij} r_{ij} + a_{ij}^0 \right),   \label{Eq:Conf}
\end{equation}
where $\boldsymbol{\lambda}_i^c$ is the usual Gell-Mann matrix of the color SU(3) group, and $a_{ij}$ and $a_{ij}^0$ are model parameters that describe the confinement strength and zero-point energy, respectively. 

The OGE potential $V_{ij}^{\text{OGE}}$ describes the short-range perturbative QCD effects. As usual, it can be written as
\begin{align}
V_{ij}^{\rm OGE} = & \; \frac{g_i g_j}{4} \boldsymbol{\lambda}_i^c \cdot \boldsymbol{\lambda}_j^c  \left[ \frac{1}{r_{ij}} - \frac{\mu_{ij}^3}{2} \frac{e^{-\mu_{ij}^2 r_{ij}^2}}{\mu_{ij} r_{ij}} \left( \frac{1}{m_i^2} + \frac{1}{m_j^2} \right. \right. \notag \\[3pt]
& \left.\left. + \, \frac{4}{3} \frac{\boldsymbol{\sigma}_i \cdot \boldsymbol{\sigma}_j}{m_i m_j}  \right)  \right],   \label{Eq:OGE}
\end{align}
where $g_{i(j)}$ is the OGE coupling constant for the $i(j)$-th constituent quark, and $\mu_{ij}$ is defined as $\mu_{ij}\equiv \beta \frac{m_i m_j}{m_i+m_j}$ with $\beta$ being a model parameter. 

Note that in Eqs.~(\ref{Eq:Conf})-(\ref{Eq:OGE}), the Gell-Mann matrix $\boldsymbol{\lambda}^c$ for a quark should be replaced by $-{\boldsymbol{\lambda}^c}^*$ for an antiquark.

In the chiral SU(3) quark model, the GBEs potential is introduced in such a way that the Lagrangian of the  quark and chiral fields is invariant under the chiral SU(3) transition \cite{Zhang:1997ny,Huang:2004ke,Huang:2004sj,Huang:2018rpb}, which gives a natural explanation of the relatively large constituent quark masses via the mechanism of spontaneous chiral symmetry breaking. At the meanwhile, the Goldstone bosons obtain their physical masses via the obvious chiral symmetry breaking caused by the tiny current quark masses. The GBEs potential provides the necessary medium- and long-range attraction in light quark systems. In the $NN$ systems, it has been shown that such attraction is rather important for describing the experimental data \cite{Huang:2018rpb}. 

The GBEs potential between a pair of light quarks reads 
\begin{equation}
V_{ij}^{\text{GBE}} = \sum_{a=0}^8 V_{ij}^{\sigma_a} + \sum_{a=0}^8 V_{ij}^{\pi_a},
\end{equation}
where the first and second terms represent the potential stemming from the exchanges of the scalar nonet mesons and pseudoscalar nonet mesons, respectively. The explicit expressions of $V_{ij}^{\sigma_a}$ and $V_{ij}^{\pi_a}$ are
\begin{align}
V_{ij}^{\sigma_a} = & - C(g_{\mathrm{ch}}, m'_{\sigma_a}, \Lambda) \, Y_1(m'_{\sigma_a}, \Lambda, r_{ij}) \left(\lambda_i^a \lambda_j^a\right), \\[6pt]
V_{ij}^{\pi_a} = & ~ C(g_{\mathrm{ch}}, m'_{\pi_a}, \Lambda) \frac{m_{\pi_a}^{\prime 2} p_{ij}^a}{48} Y_3(m'_{\pi_a}, \Lambda, r_{ij}) \left(\lambda_i^a \lambda_j^a \right) \notag \\[3pt]
& \times \left( \boldsymbol{\sigma}_i \cdot \boldsymbol{\sigma}_j \right),  
\end{align}
where
\begin{align}
C(g_{\mathrm{ch}}, m, \Lambda) = & ~ \frac{g_{\mathrm{ch}}^2}{4 \pi} \frac{\Lambda^2}{\Lambda^2-m^2} m,  \\[6pt]
Y_1(m, \Lambda, r) = & ~ Y(m r) - \frac{\Lambda}{m} Y(\Lambda r), \\[6pt]
Y_3(m, \Lambda, r) = & ~ Y(m r) - \left(\frac{\Lambda}{m}\right)^3 Y(\Lambda r), \\[6pt]
Y(x) = & ~ \frac{1}{x} e^{-x},
\end{align}
with 
\begin{align}
p_{ij}^a = & \left\{ \begin{array}{lcl} \dfrac{4}{m_i m_j}, && (a=0,1,2,3,8) \\[9pt] \dfrac{\left(m_i+m_j\right)^2}{m_i^2 m_j^2}, && (a=4,5,6,7) \end{array} \right.  \\[6pt]
m_{\sigma_a}^\prime = & \left\{ \begin{array}{ll} m_{\sigma_a}, & (a=0,1,2,3,8) \\[6pt]  \sqrt{m_{\sigma_a}^2 - \left( m_i - m_j \right)^2}, & (a=4,5,6,7)  \end{array} \right. \\[6pt] 
m_{\pi_a}^\prime = & \left\{ \begin{array}{ll} m_{\pi_a}, & (a=0,1,2,3,8) \\[6pt]  \sqrt{m_{\pi_a}^2 - \left( m_i - m_j \right)^2}. & (a=4,5,6,7) \end{array} \right. 
\end{align}
Here, $g_{\rm ch}$ is the quark and chiral field coupling constant, $\Lambda$ is the cutoff parameter indicating the chiral symmetry breaking scale, and $m_{\pi_a}$ and $m_{\sigma_a}$ $(a=0,1,2,\cdots,8)$ represent the masses of nonet pseudoscalar and nonet scalar mesons, respectively. 

For pseudoscalar meson exchanges, the mixing of $\eta_0$ and $\eta_8$ is considered,
\begin{align}
\left\{ \begin{array}{l} \eta = \eta_8 \cos\theta - \eta_0 \sin\theta,  \\[6pt]  \eta^\prime = \eta_8 \sin\theta + \eta_0 \cos\theta,  \end{array} \right.
\end{align} 
with the mixing angle taken as the empirical value $\theta=-23^\circ$.

\begin{table}[tbp]
\caption{Model parameters associated with heavy quarks. The charm quark mass $m_c$ and the zero point energies $a_{cc}^0$ and $a_{cu}^0$ in confinement potential are in MeV. The strengths of confinement $a_{cc}$ and $a_{cu}$ are in MeV/fm. }
\begin{tabular*}{\columnwidth}{@{\extracolsep{\fill}}cccccc}
\hline\hline
$m_c$ & $g_c$ & $a_{cc}$ & $a_{cu}$  & $a_{cc}^0$ & $a_{cu}^0$   \\ \hline
$1500$ & $0.635$  &  $183.8$  &  $160.9$  &  $-61.9$  &  $-95.3$    \\
\hline\hline
\end{tabular*}   \label{tab:parameters}
\end{table}

\begin{table}[tbp]
\caption{Masses (in MeV) of charmed mesons and baryons calculated by use of the parameters listed in Table~\ref{tab:parameters}. The corresponding masses from PDG \cite{ParticleDataGroup:2022pth} are listed in the last column.}
\begin{tabular*}{\columnwidth}{@{\extracolsep{\fill}}lccc}
 \hline\hline
    Particles  &  $IJ^P$  &  Masses  &  PDG values  \\
    \hline
    $D^0$        &  $\frac{1}{2}0^-$  &  $1866.9$   &  $1864.84\pm0.05$   \\
    $D^{*+}$    &   $\frac{1}{2}1^-$  &  $2011.6$   & $2010.26\pm0.05$   \\ 
    $D_s^\pm$   &  $00^-$     & $1968.4$   & $1968.35\pm0.07$   \\
    $D_s^{*\pm}$    &  $01^-$   & $2133.3$   & $2112.2\pm0.4$   \\
    $\eta_c(1S)$    &  $00^-$    & $2975.9$   & $2983.9\pm0.4$   \\ 
    $J/\psi(1S)$    &  $01^-$     &  $3096.8$   & $3096.9\pm0.006$   \\
    $\eta_c(2S)$    &  $00^-$    &  $3613.4$   &  $3638\pm 1$   \\
    $J/\psi(2S)$    &  $01^-$     &  $3686.1$   &  $3686\pm 0.01$   \\
    $\Lambda_c^+$       &  $0\frac{1}{2}^+$   & $2245.4$   & $2286.46\pm 0.14$   \\
    $\Sigma_c(2455)$    &  $1\frac{1}{2}^+$   & $2445.4$   & $2453.97\pm 0.14$   \\
    $\Sigma_c(2520)$    &  $1\frac{3}{2}^+$   & $2517.6$   & $2518.41^{+ 0.22}_{-0.18}$   \\
    $\Xi_c^+$           &  $\frac{1}{2}\frac{1}{2}^+$ & $2456.0$   & $2467.71\pm 0.23$   \\
    $\Xi_c^{'+}$        &  $\frac{1}{2}\frac{1}{2}^+$ & $2566.7$   & $2578.2\pm 0.5$   \\
    $\Xi_c(2645)^+$     &  $\frac{1}{2}\frac{3}{2}^+$ & $2642.6$   & $2645.10\pm 0.30$   \\
    $\Omega_c^0$      &  $0\frac{1}{2}^+$          & $2680.8$   & $2695.2\pm 1.7$   \\
    $\Omega_c(2770)^0$  &  $0\frac{3}{2}^+$    &  $2764.3$   & $2765.9\pm 2.0$   \\
\hline\hline
\end{tabular*}     \label{tab:fit}
\end{table}

The model parameters for light quarks are taken from our previous work of Ref.~\cite{Huang:2018rpb}. 
In that work, we achieved a satisfactory description of the masses of octet and decuplet baryon ground states, the binding energy of deuteron, and the $NN$ scattering phase shifts and mixing parameters for partial waves up to total angular momentum $J=6$ in a fairly consistent way. These parameters are: $m_u = m_d =313$ MeV, $m_{\sigma'} = m_\kappa = m_\epsilon = 980$ MeV, $m_\sigma = 569$ MeV, $m_\pi = 138$ MeV, $m_K = 495$ MeV, $m_\eta = 549$ MeV, $m_{\eta'} = 957$ MeV, $\Lambda = 1100$ MeV, $a_{uu}=58.39$ MeV/fm, $a_{uu}^0=-24.53$ MeV, $g_{u}=1.079$, and $\beta=1.606$. Note that the values of some of these parameters are a little bit different from those of Ref.~\cite{Huang:2018rpb}, because the $\delta$-function in OGE potential in Ref.~\cite{Huang:2018rpb} has now been replaced by 
\begin{align}
\delta(\boldsymbol{r}_{ij}) \quad \to \quad \frac{\mu_{ij}^3}{\pi} \frac{e^{-\mu_{ij}^2 r_{ij}^2}}{\mu r_{ij}}
\end{align}
with 
\begin{align} 
\mu_{ij} = \beta \frac{m_i m_j}{m_i+m_j}
\end{align}
in the present work to avoid the problem of collapsing ground state as the $\delta$-potential is more attractive than $1/r^2$ and thus overpowers the kinetic energy $p^2/(2m)$ for a pair of scalar quarks (antiquarks). After this replacement, the energies of octet and decuplet baryon ground states, the binding energy of deuteron, and the $NN$ scattering phase shifts and mixing parameters for partial waves up to total angular momentum $J=6$ are refitted. The obtained fitting quality is almost the same as that in Ref.~\cite{Huang:2018rpb} by using the above-mentioned parameter values. 

The other model parameters are those associated with heavy charm quarks, i.e. $m_c$, $g_c$, $a_{cu}$, $a_{cc}$, $a_{cu}^0$, and $a_{cc}^0$. They are determined by fitting the masses of ground charmed mesons and baryons. The values of these parameters are listed in Table~\ref{tab:parameters}. The predicted masses of charmed mesons and baryons are listed in Table~\ref{tab:fit}, where the corresponding values from Particle Data Group (PDG) \cite{ParticleDataGroup:2022pth} are also listed for comparison. One sees that our calculated masses of charmed mesons and baryons are rather close to the experimental values.

\section{Results And Discussion}\label{sec:results}

\begin{figure}[tbp]
\centering
\includegraphics[width=0.49\textwidth]{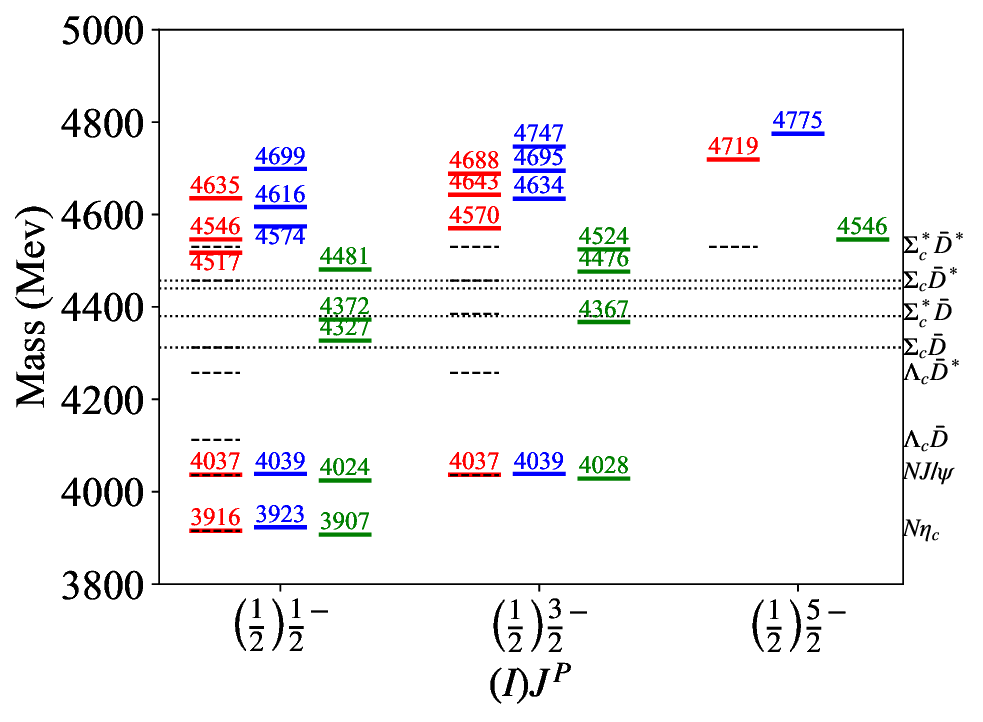}\label{Sfig:I1}
\includegraphics[width=0.49\textwidth]{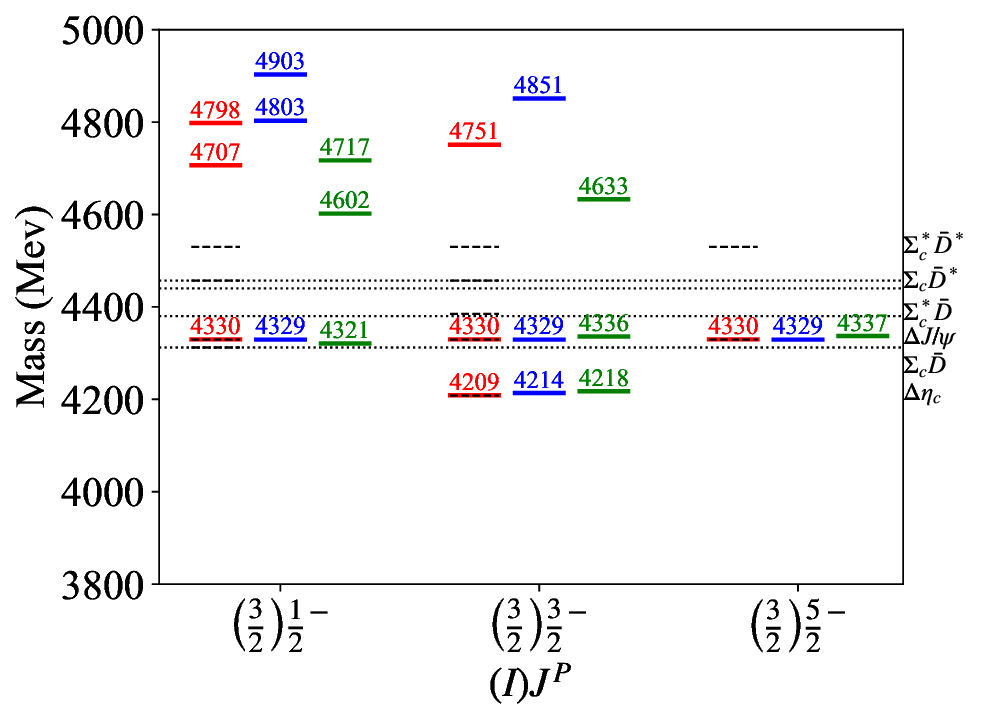}\label{Sfig:I3}
\caption{\label{Fig:mass} The mass spectra of $qqqc\bar{c}$ $(q=u,d)$ pentaquark states. The solid red, blue, green lines represent the spectra obtained in the chiral SU(3) model, the OGE model, and the CMI model, respectively. The dashed lines represent the corresponding baryon-meson thresholds, while the dotted lines represent the masses of the $P_c(4312)$, $P_c(4380)$, $P_c(4440)$, and $P_c(4457)$ states. }
\end{figure}

It is crucial to use reliable interactions between (anti)quark-(anti)quark when calculating the masses of multiquark states within a quark model. In the present work, we study the mass spectrum of $qqqc\bar{c}$ $(q = u,d)$ pentaquark systems in the chiral SU(3) quark model, where the Hamiltonian includes the kinetic energies, the OGE potential, the phenomenological confinement potential, and the GBE potentials derived from the couplings of light quarks and chiral fields. The model parameters related to light quark pairs $qq$ $(q=u,d)$ are taken from our previous work of Ref.~\cite{Huang:2018rpb}, which has been shown to be quite successful in reproducing the energies of octet and decuplet baryon ground states, the binding energy of deuteron, the $NN$ scattering phase shifts and mixing parameters for partial waves with total angular momentum up to $J=6$ in a fairly consistent way. Besides, the interactions associated with charm quark and antiquark are determined by a good fit to the masses of charmed mesons and baryons, as listed in Table~\ref{tab:fit}. We choose Gaussian functions as the trial wave functions in coordinate space and solve the Schr\"odinger equation by the variational method. The results are presented in Fig.~\ref{Fig:mass} and listed in Table~\ref{tab:hyb}.

\begin{table*}[tbp]
\caption{Predicted mass spectra, eigenvectors, and root mean square radius of $qqqc\bar{c}$ $(q=u,d)$ systems in the chiral SU(3) quark model. The masses are in units of MeV and root mean square radius are in units of fm. The $r_{qq}$ and $r_{c\bar{c}}$ represent the distances between light quarks and charm quarks, respectively. $\xi_4$ defined in Eq.~(\ref{Eq:Jacobi-xi4}) represents the distance between the center of the $qqq$ cluster and the center of the $c\bar{c}$ cluster, and $r$ defined in Eq.~(\ref{eq:r}) represents the root mean square radius of the entire pentaquark system. }
\begin{tabular*}{\textwidth}{@{\extracolsep\fill}cccccccc}
\hline\hline
   $IJ^P$ & Configuration & Mass & Eigenvector & $\sqrt{\left\langle r^2_{qq}\right\rangle}$ & $\sqrt{\left\langle r^2_{c\bar{c}}\right\rangle}$ & $\sqrt{\big\langle \xi_4^2\big\rangle}$ & $\sqrt{\left\langle r^2 \right\rangle}$  \\  \hline 
$\frac{1}{2}{\frac{1}{2}}^-$  &  $\left\{ \left(qqq\right)^{{\bm{1}}_c}_{1/2}\left[c\bar{c}\right]^{{\bm{1}}_c}_{0}\right\}^{{\bm{1}}_c}_{1/2}$  & \multirow{5}{*}{  $ \left[  \begin{array}{c}  3916 \\[8pt] 4037 \\[8pt] 4517 \\[8pt] 4546 \\[8pt] 4635  \end{array} \right] $ }   &  \multirow{5}{*}{  $ \left[ \begin{array}{rrrrr} 1 & 0 & 0 & 0 & 0 \\[8pt]  0 & 1 & 0 & 0 & 0 \\[8pt]  0 & 0 & -0.637 & 0.300 & -0.709 \\[8pt] 0 & 0 & -0.543 & 0.476 & 0.690 \\[8pt] 0 & 0 & 0.545 & 0.826 & -0.140 \end{array}  \right] $ }  & $0.813$  &  $0.363$ & $\infty$ & $\infty$  \\[6pt]
 & $\left\{\left(qqq\right)^{{\bm{1}}_c}_{1/2}\left[c\bar{c}\right]^{{\bm{1}}_c}_{1}\right\}^{{\bm{1}}_c}_{1/2}$  & & & $0.813$ & $0.450$ & $\infty$ & $\infty$ \\[6pt]
 & $\left\{\left(qqq\right)^{{\bm{8}}_c}_{3/2}\left[c\bar{c}\right]^{{\bm{8}}_c}_{1}\right\}^{{\bm{1}}_c}_{1/2}$  & & & $0.989$ & $0.712$ & $0.465$ & $0.571$ \\[6pt]
 & $\left\{\left(qqq\right)^{{\bm{8}}_c}_{1/2}\left[c\bar{c}\right]^{{\bm{8}}_c}_{0}\right\}^{{\bm{1}}_c}_{1/2}$ & & & $0.977$  & $0.715$ & $0.466$ & $0.568$ \\[6pt]
 & $\left\{\left(qqq\right)^{{\bm{8}}_c}_{1/2}\left[c\bar{c}\right]^{{\bm{8}}_c}_{1}\right\}^{{\bm{1}}_c}_{1/2}$   & & & $0.995$ & $0.715$ & $0.465$ & $0.574$ \\[9pt]
$\frac{1}{2}{\frac{3}{2}}^-$  &   $\left\{\left(qqq\right)^{{\bm{1}}_c}_{1/2}\left[c\bar{c}\right]^{{\bm{1}}_c}_{1}\right\}^{{\bm{1}}_c}_{3/2}$  & \multirow{4}{*}{  $ \left[ \begin{array}{c}  4037 \\[8pt] 4570 \\[8pt] 4643 \\[8pt] 4688  \end{array} \right] $ }   &  \multirow{4}{*}{ $ \left[ \begin{array}{rrrr} 1 & 0 & 0 & 0 \\[8pt]  0 & -0.305 & -0.341 & -0.888 \\[8pt]  0 & -0.385 & 0.897 & -0.212 \\[8pt]  0 & 0.870 & 0.277 & -0.400 \end{array}  \right] $ }  & $0.813$  & $0.450$  & $\infty$  & $\infty$  \\[6pt] 
 & $\left\{\left(qqq\right)^{{\bm{8}}_c}_{3/2}\left[c\bar{c}\right]^{{\bm{8}}_c}_{0}\right\}^{{\bm{1}}_c}_{3/2}$ &&&  $1.002$ & $0.733$  & $0.476$ & $0.582$ \\[6pt]
 & $\left\{\left(qqq\right)^{{\bm{8}}_c}_{3/2}\left[c\bar{c}\right]^{{\bm{8}}_c}_{1}\right\}^{{\bm{1}}_c}_{3/2}$ &&& $1.026$  & $0.723$  & $0.471$ & $0.588$ \\[6pt]
 & $\left\{\left(qqq\right)^{{\bm{8}}_c}_{1/2}\left[c\bar{c}\right]^{{\bm{8}}_c}_{1}\right\}^{{\bm{1}}_c}_{3/2}$ &&& $1.034$  & $0.739$  & $0.477$ & $0.594$ \\[9pt]
$\frac{1}{2}{\frac{5}{2}}^-$  &  $\left\{\left(qqq\right)^{{\bm{8}}_c}_{3/2}\left[c\bar{c}\right]^{{\bm{8}}_c}_{1}\right\}^{{\bm{1}}_c}_{5/2}$  & $4719$ & $1$ & $1.062$ & $0.756$ & $0.488$ & $0.609$  \\[9pt]
$\frac{3}{2}{\frac{1}{2}}^-$  &  $\left\{\left\{qqq\right\}^{{\bm{1}}_c}_{3/2}\left[c\bar{c}\right]^{{\bm{1}}_c}_{1}\right\}^{{\bm{1}}_c}_{1/2}$  &  \multirow{3}{*}{  $ \left[ \begin{array}{c}  4330 \\[7pt] 4707 \\[7pt] 4798  \end{array} \right] $ }   &  \multirow{3}{*}{  $ \left[ \begin{array}{rrr} 1 & 0 & 0 \\[7pt]  0 & 0.695 & 0.718 \\[7pt]  0 & -0.718 & 0.695 \end{array} \right] $ }  &  $1.171$ & $0.450$ & $\infty$ & $\infty$  \\[6pt]
 & $\left\{\left\{qqq\right\}^{{\bm{8}}_c}_{1/2}\left[c\bar{c}\right]^{{\bm{8}}_c}_{0}\right\}^{{\bm{1}}_c}_{1/2}$ &&& $1.103$ & $0.745$ & $0.481$ & $0.620$ \\[6pt]
 & $\left\{\left\{qqq\right\}^{{\bm{8}}_c}_{1/2}\left[c\bar{c}\right]^{{\bm{8}}_c}_{1}\right\}^{{\bm{1}}_c}_{1/2}$ &&& $1.103$  & $0.745$ & $0.481$ & $0.620$ \\[9pt]
$\frac{3}{2}{\frac{3}{2}}^-$  &  $\left\{\left\{qqq\right\}^{{\bm{1}}_c}_{3/2}\left[c\bar{c}\right]^{{\bm{1}}_c}_{0}\right\}^{{\bm{1}}_c}_{3/2}$  & \multirow{3}{*}{  $ \left[ \begin{array}{c}  4209 \\[7pt] 4330 \\[7pt] 4751 \end{array} \right] $ }   &  \multirow{3}{*}{  $ \left[  \begin{array}{rrr}  1 & 0 & 0 \\[7pt]  0 & 1 & 0 \\[7pt]  0 & 0 & 1  \end{array}  \right] $ }  & $1.171$ & $0.363$ & $\infty$ & $\infty$ \\[6pt]
 & $\left\{\left\{qqq\right\}^{{\bm{1}}_c}_{3/2}\left[c\bar{c}\right]^{{\bm{1}}_c}_{1}\right\}^{{\bm{1}}_c}_{3/2}$ &&& $1.171$ & $0.450$ & $\infty$ & $\infty$ \\[6pt]
 & $\left\{\left\{qqq\right\}^{{\bm{8}}_c}_{1/2}\left[c\bar{c}\right]^{{\bm{8}}_c}_{1}\right\}^{{\bm{1}}_c}_{3/2}$ &&& $1.102$ & $0.742$ & $0.481$ & $0.619$ \\[9pt]
$\frac{3}{2}{\frac{5}{2}}^-$  &  $\left\{\left\{qqq\right\}^{{\bm{1}}_c}_{3/2}\left[c\bar{c}\right]^{{\bm{1}}_c}_{1}\right\}^{{\bm{1}}_c}_{5/2}$  & $4330$ & $1$ & $1.171$ & $0.450$ & $\infty$ & $\infty$ \\
\hline\hline
\end{tabular*}     \label{tab:hyb}
\end{table*}

In Fig.~\ref{Fig:mass}, the red lines represent the results calculated in our chiral SU(3) quark model, and the dashed lines represent the thresholds of corresponding baryon-meson channels that can couple to the $qqqc\bar{c}$ $(q = u,d)$ pentaquark states as they have the same quantum numbers. The dotted lines represent the experimentally observed $P_c(4312)$, $P_c(4380)$, $P_c(4440)$, and $P_c(4457)$ states. 

One sees from Table~\ref{tab:hyb} that the following states have the color configuration $\left | {\bf{1}}_c {\bf{1}}_c \right \rangle$ [cf. Eq.~(\ref{eq:color-1c1c})]: the states $\left\{ \left(qqq\right)^{{\bm{1}}_c}_{1/2}\left[c\bar{c}\right]^{{\bm{1}}_c}_{0}\right\}^{{\bm{1}}_c}_{1/2}$ and $\left\{\left(qqq\right)^{{\bm{1}}_c}_{1/2}\left[c\bar{c}\right]^{{\bm{1}}_c}_{1}\right\}^{{\bm{1}}_c}_{1/2}$ with $IJ^P=\frac{1}{2}{\frac{1}{2}}^-$, the state $\left\{\left(qqq\right)^{{\bm{1}}_c}_{1/2}\left[c\bar{c}\right]^{{\bm{1}}_c}_{1}\right\}^{{\bm{1}}_c}_{3/2}$ with $IJ^P=\frac{1}{2}{\frac{3}{2}}^-$, the state $\left\{\left\{qqq\right\}^{{\bm{1}}_c}_{3/2}\left[c\bar{c}\right]^{{\bm{1}}_c}_{1}\right\}^{{\bm{1}}_c}_{1/2}$ with $IJ^P=\frac{3}{2}{\frac{1}{2}}^-$, the states $\left\{\left\{qqq\right\}^{{\bm{1}}_c}_{3/2}\left[c\bar{c}\right]^{{\bm{1}}_c}_{0}\right\}^{{\bm{1}}_c}_{3/2}$ and $\left\{\left\{qqq\right\}^{{\bm{1}}_c}_{3/2}\left[c\bar{c}\right]^{{\bm{1}}_c}_{1}\right\}^{{\bm{1}}_c}_{3/2}$ with $IJ^P=\frac{3}{2}{\frac{3}{2}}^-$, and the state $\left\{\left\{qqq\right\}^{{\bm{1}}_c}_{3/2}\left[c\bar{c}\right]^{{\bm{1}}_c}_{1}\right\}^{{\bm{1}}_c}_{5/2}$ with $IJ^P=\frac{3}{2}{\frac{5}{2}}^-$. The energies of these states, as shown in Fig.~\ref{Fig:mass} and Table~\ref{tab:hyb}, are the lowest among those with the same quantum numbers. Actually, these states locate just at the corresponding baryon-meson thresholds. This is because that the lowest states given by the variational method are scattering states with color configuration $\left | {\bf{1}}_c {\bf{1}}_c \right \rangle$. As the two clusters $qqq$ and $c\bar{c}$ are both colorless, the color dependent interactions, i.e. OGE potential and confinement potential, vanish between two clusters. In addition, there is no Goldstone-boson exchanges between the clusters $qqq$ and $c\bar{c}$ in the chiral SU(3) quark model. Therefore, the lowest states with color configuration $\left | {\bf{1}}_c {\bf{1}}_c \right \rangle$ tend to be free baryon-meson states and their energies are at the corresponding baryon-meson thresholds. 

In Table~\ref{tab:hyb}, we also show in the last four columns the root mean squares of $r_{qq}$, $r_{c\bar{c}}$, $\xi_4$, and $r$, which represent the distances of a light quark pair, the charm quark and antiquark pair, the centers of the $qqq$ cluster and $c\bar{c}$ cluster, and a quark to the center-of-mass of the pentaquark system. Here $\xi_4$ is defined in Eq.~(\ref{Eq:Jacobi-xi4}), and the root mean square radius $\sqrt{\langle r^2 \rangle}$ is defined as 
\begin{equation}   \label{eq:r}
\sqrt{\left\langle {\bm{r}}^2 \right\rangle} =\sqrt{ \frac{1}{5} \sum_{i=1}^{5} \left\langle  ({\bm{r}}_i - {\bm{R}}_{\rm cm})^2 \right\rangle } \, ,
\end{equation}
with ${\bm{R}}_{\rm cm}\equiv \left[ m_u \left({\bm{r}}_1 + {\bm{r}}_2 + {\bm{r}}_3\right) + m_c \left({\bm{r}}_4 + {\bm{r}}_5\right) \right] / \left(3m_u + 2m_c\right)$ being the coordinate of the center-of-mass motion of the $qqqc\bar{c}$ $(q = u,d)$ pentaquark system. One sees that for the states with color configuration $\left | {\bf{1}}_c {\bf{1}}_c \right \rangle$, the values of $\sqrt{\big\langle \xi_4^2\big\rangle}$ tend to be infinity that coincides with the fact that these states are free scattering states, locating just at the corresponding baryon-meson thresholds.
 
Note that the states with color configuration $\left | {\bf{1}}_c {\bf{1}}_c \right \rangle$ do not couple to those with color configuration $\left | {\bf{8}}_c {\bf{8}}_c \right \rangle$, as by variational method the distance between the two clusters of $qqq$ and $c\bar{c}$ in the states with color configuration $\left | {\bf{1}}_c {\bf{1}}_c \right \rangle$ tends to be infinity, which causes zero transition matrix elements in coordinate space. 

The states with color configuration $\left | {\bf{8}}_c {\bf{8}}_c \right \rangle$ have much higher masses as shown in Fig.~\ref{Fig:mass} and listed in Table~\ref{tab:hyb}. They all lie much above the baryon-meson thresholds, thus are not suggested as compact pentaquark states, as they can decay to baryon-meson channels via quark rearrangement, known as fall-apart decays. Moreover, the calculated masses of all the states with color configuration $\left | {\bf{8}}_c {\bf{8}}_c \right \rangle$ are much higher than the experimental masses of the $P_c$ states. In view of this, we conclude that the chiral SU(3) quark model does not support any narrow and compact $qqqc\bar{c}$ $(q = u,d)$ pentaquark states. In particular, the experimentally observed $P_c$ states, i.e. $P_c(4312)$, $P_c(4380)$, $P_c(4440)$, and $P_c(4457)$, can not be accommodated as compact pentaquark states in the chiral SU(3) quark model.

\begin{table}[tbp]
\caption{Contributions from individual terms of Hamiltonian in the chiral SU(3) quark model, the OGE model, and the CMI model.}
\begin{tabular*}{\columnwidth}{@{\extracolsep\fill}lrrr}
\hline\hline
  & chiral SU(3) & OGE  & CMI  \\ 
\hline
$\sum_i m_i$ & $3939$  &  $4347$  &  \\
$T$  &  $855$ & $1095$ & \\
$V^{\text{Conf}}$ & $426$ & $-5$ & \\
$V^{\text{Coul}}$ & $-769$ & $-1024$ & \\
$V^{\text{CE}}$   & $225$ & $276$ & \\
$V^{\text{CM}}$  & $77$ & $86$ & $98$ \\
$V^{\text{S}}$     & $-47$ & $0$  & \\
$V^{\text{PS}}$   & $13$ & $0$  & \\
$\sum_i m_i + V^{\text{Conf}}$       & $4365$  &  $4342$  & \\
$T + V^{\text{Coul}}$     &  $86$  & $71$  &  \\
$V^{\text{CE}} + V^{\text{CM}}$    & $302$ & $362$ & \\
$V^{\text{S}} + V^{\text{PS}}$       & $-34$ & $0$ & \\
$H$   &  $4719$  &  $4775$  & $4546$  \\
\hline\hline
\end{tabular*}  \label{Table:each_term}
\end{table}

Although the GBE potential are known to be indispensable for the medium and long range $NN$ interaction, the constituent quark model that consists only the OGE potential, namely, the OGE model, has also been commonly used in literature in studying the hadron spectroscopy. To make a comparison, we also compute the mass spectrum of the $qqqc\bar{c}$ $(q = u,d)$ systems in the OGE model with the model parameters being determined by a fit of the masses of octet and decuplet light baryons and the ground charmed mesons and baryons. The results of calculated masses of the $qqqc\bar{c}$ $(q = u,d)$ states are shown in Fig.~\ref{Fig:mass} with solid blue lines. It is seen that for the states with color configurations $\left | {\bf{1}}_c {\bf{1}}_c \right \rangle$, the masses calculated in the OGE model, similar to those in the chiral SU(3) quark model, are at the corresponding baryon-meson thresholds. Note that the baryon-meson thresholds calculated in the OGE model differ with those in the chiral SU(3) quark model by a few MeV due to the differences of the parameters fitted in these two models. The reason why the computed masses of the lowest states are at the baryon-meson thresholds is that in the OGE model these states are also scattering states as there is no interaction between the two color singlet clusters, $qqq$ and $c\bar{c}$. For other states with color configurations $\left | {\bf{8}}_c {\bf{8}}_c \right \rangle$, the masses calculated in the OGE model are always higher that those in the chiral SU(3) quark model. These high-mass states are not suggested as compact pentaquark states as they can easily decay to baryon-meson channels with the same quantum numbers via quark rearrangement. One thus concludes that the OGE model cannot accommodate the $qqqc\bar{c}$ $(q = u,d)$ compact pentaquark sates either, and the experimentally observed $P_c$ states should be interpreted by other scenarios. 

For a better understanding of the mass difference obtained in the chiral SU(3) quark model and the OGE model, in Table~\ref{Table:each_term} we list in the third column the contributions of individual terms of the Hamiltonian to the mass of the state $\left\{\left(qqq\right)^{{\bm{8}}_c}_{3/2}\left[c\bar{c}\right]^{{\bm{8}}_c}_{1}\right\}^{{\bm{1}}_c}_{5/2}$ with isospin spin-parity $IJ^P=\frac{1}{2}{\frac{5}{2}}^-$. Here $\sum_i m_i$ is the sum of constituent quark masses, and $T$, $V^{\rm Conf}$, $V^{\rm Coul}$, $V^{\rm CE}$, $V^{\rm CM}$, $V^S$, and $V^{PS}$ represent the contributions of kinetic energy, confinement potential, color coulomb potential, color electric potential, color magnetic potential, scalar meson exchange potential, and the pseudoscalar meson exchange potential, respectively. One sees that the sum of $\sum_i m_i$ and $V^{\rm Conf}$ in the OGE model is a little bit lower than that in the chiral SU(3) quark model, so does the sum of $T$ and $V^{\rm Coul}$. However, the sum of $V^{\rm CE}$ and $V^{\rm CM}$ offers about $60$ MeV more repulsion in the OGE model than in the chiral SU(3) quark model. Furthermore, the sum of $V^{\rm S}$ and $V^{\rm PS}$ offers $34$ MeV attractive in the chiral SU(3) quark model while it is absent in the OGE model. Stronger repulsion from $V^{\rm CE}+V^{\rm CM}$ and less attraction from $V^{\rm S}+V^{\rm PS}$ explains why the mass calculated in the OGE model is higher than that in the chiral SU(3) model.

In Ref.~\cite{Weng:2019ynv}, Weng {\it et al.} studied the mass spectrum of the hidden charm pentaquark states in the CMI model. In their work, apart from the scattering states with color configurations $\left | {\bf{1}}_c {\bf{1}}_c \right \rangle$, all other states with color configurations $\left | {\bf{8}}_c {\bf{8}}_c \right \rangle$ and isospin $I=1/2$ were suggested as compact pentaquark states. We have repeated their calculations and the results are shown in Fig.~\ref{Fig:mass} with solid green lines. Specifically, in their work, the state with mass $M=4327$ MeV and isospin spin-parity $IJ^P=\frac{1}{2}{\frac{1}{2}}^-$ was suggested to be the $P_c(4312)$ state, the state with $M=4367$ MeV and $IJ^P=\frac{1}{2}{\frac{3}{2}}^-$ and the state with $M=4372$ MeV and $IJ^P=\frac{1}{2}{\frac{1}{2}}^-$ were suggested as the $P_c(4380)$ state, the state with $M=4476$ MeV and $IJ^P=\frac{1}{2}{\frac{3}{2}}^-$ was suggested to be the $P_c(4440)$ state, the state with $M=4481$ MeV and $IJ^P=\frac{1}{2}{\frac{1}{2}}^-$ was suggested as the $P_c(4450)$ state, and two other states, one with $M=4525$ MeV and $IJ^P=\frac{1}{2}{\frac{3}{2}}^-$ and the other with $M=4546$ MeV and $IJ^P=\frac{1}{2}{\frac{5}{2}}^-$, were suggested as new predictions of the pentaquark states to be observed by experiments.

In our chiral SU(3) quark model, the calculated masses of all the states with color configurations $\left | {\bf{8}}_c {\bf{8}}_c \right \rangle$ are about $154\sim203$ MeV higher than those obtained in the CMI model of Ref.~\cite{Weng:2019ynv}, and are also far away from the experimental values of the masses of those $P_c$ states. Contrary to the CMI model of Ref.~\cite{Weng:2019ynv}, our chiral SU(3) quark model shows that the experimentally observed $P_c$ states cannot be accommodated as compact pentaquark states. The reason why the masses calculated in our chiral SU(3) quark model are much higher than those in the CMI model can be understood by the following analysis. 

In the CMI model of Ref.~\cite{Weng:2019ynv}, the Hamiltonian was written as 
\begin{align}
H = -\frac{3}{16} \sum_{i<j} m_{ij} {\bm{\lambda}}_i^c \cdot {\bm{\lambda}}_j^c - \frac{1}{16}\sum_{i<j} v_{ij} {\bm{\sigma}}_i \cdot {\bm{\sigma}}_j {\bm{\lambda}}_i^c \cdot {\bm{\lambda}}_j^c,    \label{eq:CMIH}
\end{align}
where $m_{ij}$ and $v_{ij}$ are model parameters fixed by a fit of the masses of traditional mesons and baryons. The second term represents the color-magnetic interaction, and the first term incorporates the contributions from all other terms, e.g. the constituent quark mass, kinetic energy, phenomenological confine interaction, color-electric interaction, and so on. In Table~\ref{Table:each_term}, we list in the forth column the individual contributions from the CMI model to the mass of the state $\left\{\left(qqq\right)^{{\bm{8}}_c}_{3/2}\left[c\bar{c}\right]^{{\bm{8}}_c}_{1}\right\}^{{\bm{1}}_c}_{5/2}$ with isospin spin-parity $IJ^P=\frac{1}{2}{\frac{5}{2}}^-$. One sees that the $V^{\text{CM}}$ is $98$ MeV in the CMI model, $21$ MeV higher than the value in the chiral SU(3) model. However, the calculated mass, i.e. the total contribution $H$, in the CMI model is $173$ MeV lower than that in the chiral SU(3) quark model. This means in the CMI model, a much bigger attraction has been absorbed into the first term of $H$ in Eq.~(\ref{eq:CMIH}). 

In the CMI model, the matrix elements in coordinate space have been parameterized as constants [see Eq.~(\ref{eq:CMIH})]. These constants are fixed by a fit of the masses of traditional mesons and baryons and then applied to the pentaquark systems. This means that the distance between two quarks in a pentaquark state is assumed to be the same as that in a traditional meson or baryon. However, in our chiral SU(3) quark model calculation, the calculated distance between two quarks, $r_{qq}$ or $r_{c\bar{c}}$, in a pentaquark quark state with color configuration $\left | {\bf{8}}_c {\bf{8}}_c \right \rangle$ is much bigger than that in the scattering sate which has color configuration $\left | {\bf{1}}_c {\bf{1}}_c \right \rangle$ and can be treated as free baryon-meson state, as shown in Table~\ref{tab:hyb}.

In Table~\ref{Table:each_term}, one sees that the dominant attraction is coming from the color coulomb interaction. As the color coulomb interaction is proportional to $1/r$, a much smaller distance between a pair of quarks means a much stronger coulomb attraction between them. This probably explains why the mass of pentaquark state calculated in the CMI model is much lower than that in the chiral SU(3) quark model. A similar analysis was also given in Ref.~\cite{Deng:2020iqw} for a study of fully-heavy tetraquark systems.

\section{Summary}\label{sec:summary}

The $P_c$ states reported by the LHCb Collaboration have aroused a lot of theoretical investigations. But up to now the structure of $P_c$ states is still an open question, as inconclusive results were obtained from different theoretical works, regardless of whether these works were completed with the same or different theoretical methods. This poor situation urges upon us further independent analysis with reliable model ingredients to get a better understanding of the structures of the hidden-charm pentaquark states. 

In our previous work of Ref.~\cite{Huang:2018rpb}, we have successfully described the energies of octet and decuplet baryon ground states, the binding energy of deuteron, the $NN$ scattering phase shifts and mixing parameters for partial waves with angular momentum up to $J=6$ in a quite consistent way within a chiral SU(3) quark model. The Hamiltonian includes the kinetic energy, the OGE potential, the confinement potential, and the GBE potential. 

In the present work, we further extend our chiral SU(3) quark model employed in Ref.~\cite{Huang:2018rpb} to explore the mass spectra of the $qqqc\bar{c}$ $(q = u,d)$ pentaquark systems. The interactions between a pair of light quarks are taken from Ref.~\cite{Huang:2018rpb}. The interactions associated with charm quark and antiquark consist of the OGE potential and the confinement potential, and the corresponding parameters are fixed by a fit to the masses of known charmed baryons and mesons. The masses and eigenvectors for the $qqqc\bar{c}$ $(q = u,d)$ pentaquark states are obtained by solving the Schr\"odinger equation via the variational method. 

Our results show that all the states with color configuration $\left | {\bf{8}}_c {\bf{8}}_c \right \rangle$ have much higher masses than the corresponding baryon-meson thresholds, thus are not suggested as compact pentaquark states, as they can decay to baryon-meson channels via quark rearrangement. Moreover, the calculated masses of all the states with color configuration $\left | {\bf{8}}_c {\bf{8}}_c \right \rangle$ are much higher than the experimental masses of the $P_c$ states. We conclude that the chiral SU(3) quark model does not support any narrow and compact $qqqc\bar{c}$ $(q = u,d)$ pentaquark states, and in particular, the experimentally observed $P_c$ states, i.e. $P_c(4312)$, $P_c(4380)$, $P_c(4440)$, and $P_c(4457)$, can not be accommodated as compact pentaquark states in the chiral SU(3) quark model.

We also calculate the masses of the $qqqc\bar{c}$ $(q = u,d)$ pentaquark systems in the OGE model, where the interactions incorporate only the OGE potential and the confinement potential. It is found that the masses of hidden-charm pentaquark states obtained in the OGE model are much higher than those in the chiral SU(3) quark model, as the OGE model provides stronger repulsion from the color electric and color magnetic interactions and missing attraction from the GBE interaction.  

In Ref.~\cite{Weng:2019ynv}, it was claimed that all the experimentally observed $P_c(4312)$, $P_c(4380)$, $P_c(4440)$, and $P_c(4457)$ states can be explained as compact pentaquark states with $I=1/2$ in the CMI quark model, as the calculated masses of pentaquark states are very close to the experimental values of the $P_c$ states. We show that the masses of $qqqc\bar{c}$ $(q = u,d)$ pentaquark states calculated in the chiral SU(3) quark model are about $154\sim 203$ MeV higher than those obtained in the CMI model, mainly because the chiral SU(3) quark model provides less color coulomb attraction as the calculated distances among quarks are much larger than those used in the CMI quark model which are assumed to be the same as those in the traditional mesons or baryons.

\begin{acknowledgments}
This work is partially supported by the National Natural Science Foundation of China under Grants No.~12175240 and No.~11635009, and the Fundamental Research Funds for the Central Universities.
\end{acknowledgments}

\bibliographystyle{apsrev}
\bibliography{qqqcc.bib}

\end{document}